\documentclass[11pt]{article}
\usepackage{latexsym}
\usepackage{amssymb}
\usepackage{amsmath}
\usepackage{amscd}
\usepackage{epsfig}
\usepackage{graphics}

\newcommand{\autor}[2]{{\it #1\,}\footnote{#2}}

\def\be#1\ee{\begin{equation}#1\end{equation}}
\def\bea#1\eea{\begin{eqnarray}#1\end{eqnarray}}

\font\openface=msbm10 at10pt
\def\SetOf#1#2{\left\{ #1  \,|\, #2 \right\} }
\def\NaturalNumbers{{\hbox{\openface N}}}

\newcommand{\Ctilde}{\tilde{c}}

\newcommand{\kmion}{k\! -\!\!\, 1}
\newcommand{\kpion}{k\! +\!\!\, 1}

\newcommand{\cS}{\mathcal{S}}
\newcommand{\tOm}{\tilde{\Omega}}
\newcommand{\Omegatilde}{\tilde{\Omega}} 
\newcommand{\Om}{\Omega}
\newcommand{\tcR}{\tilde{\mathcal{R}}}
\newcommand{\Rtilde}{\tilde{\mathcal{R}}}
\newcommand{\cR}{\mathcal{R}}
\newcommand{\R}{\mathcal{R}}

\newcommand{\cF}{\mathcal{F}}

\newcommand{\cC}{\mathcal{C}}

\newcommand{\ti}[1]{\tilde{#1}}

\newcommand{\cyl}{{\rm cyl}}
\newcommand{\stem}{{\rm{stem}}}
\newcommand{\mutilde}{\tilde{\mu}}
\newcommand{\N}{\mathbb{N}}

\newcounter{def}

\newtheorem{theorem}{Theorem}
\newtheorem{lemma}{Lemma}
\newtheorem{proposition}{Proposition}
\newtheorem{corollary}{Corollary}

\newcommand{\bproof}{\setlength{\parindent}{0mm}{\bf Proof{~~}}}
\newcommand{\eproof}{$\Box$\setlength{\parindent}{5mm}}

\newcommand{\rar}{\rightarrow}

\newcommand{\blist}{\begin{list}{}{\setlength{\leftmargin}{4mm}
\setlength{\parindent}{0mm}\setlength{\parsep}{1mm}
\setlength{\topsep}{2mm}}}
\newcommand{\elist}{\end{list}}
\begin{document}
\begin{titlepage}
\begin{center}
{\bf\LARGE ``Observables'' in causal set cosmology}
\end{center}

\bigskip

\begin{center}
\renewcommand{\thefootnote}{\alph{footnote}}
\autor{Graham Brightwell}{Department of Mathematics, London School of
Economics, Houghton Street, London WC2A 2AE, UK}, 
\autor{H. Fay Dowker}{Department of Physics, Queen Mary,
University of London, Mile End Road, London E1 4NS, UK},
\autor{Raquel S. Garc{\'\i}a}{Blackett Lab,
Imperial College, Prince Consort Road, London SW7 2AZ, UK},
\autor{Joe Henson}{Department of Physics, Queen Mary,
University of London, Mile End Road, London E1 4NS, UK},
\autor{Rafael D. Sorkin}{Department of Physics, Syracuse
University, Syracuse, NY 13244-1130, USA} \setcounter{footnote}{0}
\end{center}
\begin{center} {\large In memory of Sonia Stanciu}
\end{center}

\vspace{3cm}

\begin{abstract} 
\noindent 
The ``generic''
family of classical sequential growth dynamics 
for causal sets \cite{Rideout:1999ub}
provides cosmological models 
of causal sets which are a  
testing ground for 
ideas about the, 
as yet unknown, quantum theory. 
In particular we can investigate how general covariance 
manifests itself and address the problem of identifying and interpreting 
covariant ``observables'' in quantum gravity. 
The problem becomes, in this setting,  
that of identifying measurable covariant collections of causal sets,
to each of which corresponds the question:
``Does the causal set that occurs belong to this collection?''
It has for answer the probability measure of the collection. 
Answerable covariant questions, then, 
correspond to measurable
collections of causal sets which are independent of the
labelings 
of the causal sets. 
However, 
what
the transition probabilities of the classical sequential 
growth dynamics provide directly is 
a measure on the space of {\it labeled}
causal sets and the physical interpretation of the covariant measurable 
collections is consequently obscured.  We  show that there is a physically 
meaningful characterisation of the class of measurable covariant 
sets
as unions and differences of ``stem sets''.
\end{abstract}

\end{titlepage}

\section{Introduction}

What are the observables for quantum gravity? The question is 
often posed but its meaning is clouded by a number of 
difficulties.
One problem is that we do not even know what the 
observables are in a theory as familiar as flat spacetime 
non-abelian gauge theory \cite{Beckman:2001ck}.
 Another major problem is that the word ``observable''
is inherited from an interpretation of quantum mechanics --- the 
standard interpretation --- in which the subject matter of the theory is
not what $\it is$ but what can be $\it observed$. However adequate this
may be for laboratory science, it will not do for quantum gravity, 
and we should rather be seeking 
what Bell called 
the ``be-ables''.
Furthermore, the question is intimately tied to
the issue of general covariance, and indeed to 
the meaning of general covariance itself. It seems that 
the requirement of general covariance threatens to obscure
the physical interpretation of the theory since objects identified 
mathematically as ``covariant'' may not look like anything 
useful for making predictions.  

One advantage of the causal set approach to quantum gravity
\cite{Bombelli:1987aa, Sorkin:1990bh, Sorkin:1990bj} is that 
it is straightforward enough conceptually that we can 
address these knotty problems in a productive way. 
Although we do not yet have a quantum dynamics for causal sets
we do have a family of classical stochastic dynamics \cite{Rideout:1999ub}
within which we can investigate issues such as general covariance
and the identification of observables
completely concretely. The discreteness of
causal sets  turns out to eliminate many of the technical
difficulties that tend to obscure these issues in the
continuum. 
The work described in the current paper 
is a continuation of that reported in 
\cite{Brightwell:2002yu},  
and the main result is a proof of a conjecture made in 
that paper.  

In the next section we summarise the classical sequential growth
dynamics for causal sets.  In the context of this dynamics, the question
we started with, ``What are the observables?''  is replaced by, ``What
are the physical questions to which the dynamics provides answers?''.
We see that the dynamics provides a probability measure $\mu$ on the
sample space of all labeled causal sets as possible histories of the
universe (these are cosmological models).  Questions that we may ask are
of the form: ``Does the causal set that actually occurs --- {\it i.e.}
the real one --- belong to subset $A$ of the sample space?''  where $A$
is a $\mu$-measurable set, and the measure $\mu$ provides the answer:
``Yes, with probability $\mu(A)$.''  To be generally covariant, the
questions, {\it i.e.} the subsets of the set of all labeled causal sets,
must be independent of labeling.  Thus we are led to the identification
of the {\it covariant} questions as subsets $A$ of the set of all causal
sets such that if a certain labeled causal set is an element of $A$, so
are all its re-labelings.

This identification is, however, very abstract and what we are seeking,
then, is a characterisation of the measurable sets that will be
physically useful. In section 3 we describe the result that we will
prove, namely that 
any measurable set of causal sets can be 
formed by countable set operations on the so-called ``stem sets,''
to be defined. 
These stem sets have an accessible physical meaning.
Sections 4, 5 and 6 are devoted to proving the theorem and 
the last section contains a discussion.

\section{Classical sequential growth models and the covariant questions}

The causal set hypothesis is that the continuum spacetime of
general relativity is an approximation to a deeper level of discrete
structure which is a 
past finite 
partial order or {\em causal set (causet)}.  
This is a set endowed with a binary relation $\prec$ such that 
$(x\prec y)\, {\rm and}\, (y\prec z) \implies (x\prec z)$ (transitivity),
$x \not\prec x$ (acyclicity), 
and all ``past-sets'' $\SetOf{x}{x\preceq z}$
are finite.  (The condition that all past-sets are finite implies 
that the partial order is locally finite.  In other contexts 
one would weaken 
the condition of past finiteness to local
finiteness in the definition of a causet, but for present purposes there
is no harm in using the stronger condition.) When $x \prec y$ we say that 
``$x$ is below $y$'' or ``$y$ is above $x$.'' 
We will be interested 
in both finite and countably infinite causets.  

Although we don't yet have a 
quantum dynamics for causal sets, the generic family of classical sequential  
growth (CSG) dynamics  derived in \cite{Rideout:1999ub}  
is a good place to begin the search for physical questions, 
as a warm-up for the quantum theory when we have it. 

\begin{figure}[ht]
\centering \resizebox{!}{1.5in}{\includegraphics{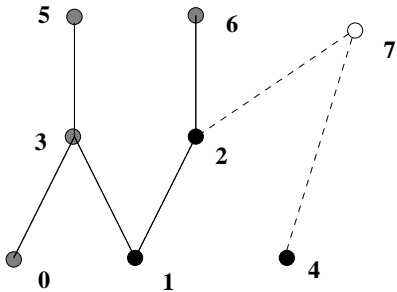}}
\caption{\small{Transition from stage 7 to stage 8 
of a particular growth process. Two vertices joined by a line 
are related, with the lower being to the past of (an 
ancestor of) the upper one. Only covering
relations (``links'')
are represented, the rest being implied by transitivity.
Points filled with black are to the past of the new element (8) and
points filled with grey are spacelike to it.} \label{growth.figure}}
\end{figure}
\vspace{0.5cm}

Each of the dynamical laws in question describes a stochastic birth
process in which elements are ``born'' one by one so that, at stage $n$,
it has produced a causet $\Ctilde_n$ of $n$ elements, within which the
most recently born element is maximal.  If one employs a genealogical
language in which ``$x\prec{y}$'' can be read as ``$x$ is an ancestor of
$y$'', then the $n^{th}$ element (counting from $0$) must at birth
``choose'' its ancestors from the elements of $\tilde{c}_n$, and for
consistency it must choose a subset $s$ with the property that
$x\prec{y}\in{s}{\,\implies\,}x{\in}s$.  (Every ancestor of one of my
ancestors is also my ancestor.)  
Such a subset $s$ (which is necessarily finite) 
will be called a {\it stem}.\footnote
{In \cite{Rideout:1999ub} this was called a ``partial stem'', but we
 will not need to draw a distinction between partial and full stems
 here.  Notice that a stem is by definition finite.  Dropping this
 finiteness requirement, we get the notion of ``down-set'' or ``past-set'', also
 called an ``ideal''.}
The dynamics is then determined fully by
giving the {\it transition probabilities} governing each such choice of
$s\subset\Ctilde_n$.

We can formalize this scheme by introducing for each integer
$n=0,1,2,\cdots$ the set $\Omegatilde(n)$ of {\it labeled} causets 
$\Ctilde_n$ whose elements are labeled by
integers $0,1,\cdots,n-1$ that record their order of birth.  Moreover
this labeling $L$ is {\it natural} in the sense that
$x{\prec}y{\,\implies\,}L(x)<L(y)$.  Each
birth of a new element 
occasions one of the allowed transitions from $\Omegatilde(n)$ to
$\Omegatilde(n+1)$ and occurs with a specified conditional probability
$\tau$.

A specific stochastic dynamics is fixed by giving the $\tau$
for all possible transitions. Under the physically motivated
assumptions of ``discrete general covariance'' and ``Bell 
Causality'' 
the possibilities for the $\tau$ are severely narrowed down and 
have been 
largely
classified in
\cite{Rideout:1999ub}. The main conclusion is 
that $\tau$ generically takes the
form 
\begin{equation}\label{transprob.eq}
\tau=\frac{\lambda(\varpi,m)}{\lambda(n,0)}
\end{equation}  
where, for the potential
transition in question, $\varpi$ is the number of ancestors of the new
element, $m$ the number of its ``parents'', and $n$ the number of
elements present before the birth, and where $\lambda(\varpi,m)$ is
given by the formula $\sum_k{\varpi-m\choose k-m}t_k$ with the 
non-negative real numbers $t_k$
being the free parameters or ``coupling constants'' of the theory.  

An alternative intepretation of this dynamical rule is as follows.  Each
new element chooses some set $s$ of elements from among those already
present, a set $s$ being chosen with relative probability $t_{|s|}$:
then the new element is placed above $s$ and all ancestors of members of
$s$.  Setting $t_k = t^k$, where $t$ is constant, gives the model of
``random graph orders'' as studied in \cite{Barak:1984}, \cite{Albert:1989},
\cite{Alon:1994} and 
\cite{Bollobas:1997}, for instance.  In \cite{Rideout:1999ub} this special
case was called ``percolation''.  One can thus understand the dynamical
rule given by (\ref{transprob.eq}) as defining a type of ``generalized
percolation model''.  These models exhaust the ``generic'' solution
family of \cite{Rideout:1999ub}, and they are the only ones we consider
in this paper.  (We thereby ignore such ``exceptional'' CSG models as
``originary percolation'', not to mention those exceptional solutions
which are not even limits of the generalized percolation form.)

There are two simple cases that
we can describe completely: if all the $t_k$ are zero except $t_0$, 
then we obtain a causet in which no pair of elements is related, while 
if only $t_0$ and $t_1$ are non-zero we almost surely obtain a 
causet which is an infinite union of trees in which every element has 
infinitely many children.  For the present, we 
rule out these two cases, so we assume that $t_k > 0$ for some $k \ge 2$, 
but we do not need to make any other assumptions regarding the coupling 
constants in the model.

The stochastic dynamics described above gives rise to a notion of 
probability that is too rich for our purposes, as it assigns a probability 
as an answer to the question ``is element~3 above element~1?'', which is 
not meaningful for us, as in any particular causet the answer depends on 
the labeling of the elements.  
In order to arrive at a definite theory one needs to specify 
the set of questions that the theory should answer and for each one
of them, explain how in principle, the  probability  of the answer 
``yes'' can be computed.

We will need to proceed in a formal manner.  We wish to construct 
a {\em probability space}, which is a 
triad consisting of: a {\it sample space} $\Omega$, 
a {\it $\sigma$-algebra} $\R$ on $\Omega$,
and a {\it probability measure} $\mu$ with domain $\R$.
In relation to the two tasks above, each member $Q$ of $\R$ corresponds to
one of the answerable questions and its measure $p=\mu(Q)$ is the
answer.  
(That $\R$ is a $\sigma$-algebra on $\Omega$ means that it is
 a non-empty family of subsets of $\Omega$ closed under complementation and
 countable intersection.  
 A probability measure $\mu$ with domain $\R$ is a function that 
 assigns to each member of $\R$ a non-negative real number --- its
 probability --- 
 such that $\mu$ is countably additive, with $\mu(\Omega)=1$.  
 Finally, countable additivity means that $\mu(\cup_n A_n) = \sum_n \mu(A_n)$ 
 for any countable collection of mutually disjoint sets in $\R$.)

In the case at hand, the sample space is the set
$\Omegatilde\equiv\Omegatilde(\infty)$ of {\it completed labeled causets}, 
these being the infinite causets that would result if the birth process
were made to ``run to completion''.  (We use a tilde to indicate
labeling.)  
The dynamics is then given by a
probability measure $\mutilde$, constructed from the transition
probabilities $\tau$, whose domain
$\Rtilde$ is a $\sigma$-algebra specified as follows. 
To each finite causet $\tilde{b}\in\Omegatilde(n)$ one can associate the
``cylinder set'' $\cyl(\tilde{b})$ 
comprising all those $\Ctilde\in\Omegatilde$
whose first $n$ elements (those labeled $0\cdots{n-1}$) form an
isomorphic copy of $\tilde{b}$ (with the same labeling); 
and $\Rtilde$ is then the smallest
$\sigma$-algebra containing all these cylinder sets.
More constructively, $\Rtilde$ is the collection of all subsets of
$\Omegatilde$ which can be built up from the cylinder sets by a
countable process involving countable union, intersection and
complementation.  The transition probabilities $\tau$ provide us with 
the probability of each cylinder set $\cyl(\tilde{b})$, and standard
results in probability theory imply that this extends to a 
probability measure on $\Rtilde$.   

For future use, we will need 
in addition to $\Omegatilde$
the corresponding space $\Omega$ of
completed {\it unlabeled} causets, whose members can also be viewed in
an obvious manner as equivalence classes within $\Omegatilde$.
We will also need the set of
all finite labeled causets 
$\tilde\Omega(\N) \equiv \cup_{n\in\N} \Omegatilde (n)$
and its unlabeled counterpart $\Omega(\N) = \cup_{n\in\N} \Omega(n)$.

At first hearing, calling a probability measure a dynamical law might
sound strange, but in fact, once we have the measure $\mutilde$ we can
say everything of a predictive nature that it is possible to say {\it a
priori} about the behavior of the causet $\Ctilde$.  For example, one might 
ask ``Will the universe recollapse?'' 
This can be interpreted mathematically as asking whether $\Ctilde$ will 
develop a ``post'',
defined as an element whose ancestors and descendants taken together
exhaust the remainder of $\Ctilde$.  Let $A\subset\Omegatilde$ be the 
set of all completed labeled causets having posts.
(One can show that $A\in\Rtilde$, so that $\mutilde(A)$ is defined.)
Then our question is equivalent to asking whether $\Ctilde\in{A}$, and the
answer is ``yes with probability $\mutilde(A)$.''    
It is thus $\mutilde$ that expresses the ``laws of motion'' (or better
``laws of growth'') that constitute our stochastic dynamics: its domain
$\Rtilde$ tells us which questions the laws can answer, and its values
$\mutilde(A)$ tell us what the answers are.

In this context, we  can see what the expression of general covariance is. 
In a causet, only the relations between elements have physical significance:
the labels on causet elements are considered as physically meaningless.
Thus, for a subset of $A\subset\Omegatilde$ to be
covariant, it cannot contain any labeled completed causet $\Ctilde$
without containing at the same time all those $\Ctilde'$ isomorphic to
$\Ctilde$ (i.e. differing only in their labelings).  To be measurable as
well as covariant, $A$ must also belong to $\Rtilde$.
Let $\R$ be the collection of all such sets:
$A\in\R{\iff} {A}\in\Rtilde
\ {\rm and}\
\forall \Ctilde_1 \simeq \Ctilde_2\in\Omegatilde,
\Ctilde_1\in{A}\implies\Ctilde_2\in{A}$.
It is not hard to see that $\R$ is a sub-$\sigma$-algebra of $\Rtilde$,
whence the restriction of $\mutilde$ to $\R$ is a measure $\mu$ on the
space $\Omega$ of unlabeled completed causets. (As 
just defined, an element $A \in \R$ is a subset of $\Omegatilde$.
However, because it is re-labeling invariant, it can also 
be regarded as a subset of $\Omega$.) Any element of 
$\R$ corresponds to a covariant question to which the 
dynamics provides the answer in the form of $\mu$. 

However, the definition of $\R$ provides no useful 
information about the physical meaning of these
covariant questions. All we know is that an element 
of $\R$ is formed from the (non-covariant) cylinder sets by doing countably
many set operations after which the resulting set must contain 
all re-labelings of each of its elements. 
Our purpose now is to provide a construction of $\R$ that is
physically useful. 

\section{The physical questions}

Among the questions belonging to $\R$ there are some which do have a
clear significance.  
Let $b\in\Omega(\N)$ be a finite unlabeled causet and let
$\stem(b)\subset\Omega$ be the ``stem set'',
$\stem(b) = \SetOf{c\in\Omega}{c \mbox{ contains a stem isomorphic to } b}$.
Thus $\stem(b)$ comprises those unlabeled completed causets with the
property that there exists a natural labeling such that the first $n$
elements form a causet isomorphic to $b$.  
Each stem set 
$\stem(b)$ --- treated as a subset of $\Omegatilde$ --- is 
a countable union of cylinder sets:
$$
  \stem(b)=
  \bigcup
  \SetOf
  {\cyl(\Ctilde)}
  {\Ctilde\in\Omega(\NaturalNumbers) 
   \mbox{ and } b \mbox{ is a stem in } 
   \Ctilde }
$$
Therefore the stem sets belong to $\Rtilde$ and hence to $\R$.  
For this particular element of $\R$, the meaning of the corresponding
causet question is evident: 
``Does the causet contain $b$ as a stem?''.\footnote
{Strictly,
``$a$ is a stem  in $c$'' and ``$c$ contains $a$ as a stem''
mean that $a$ is a subset of $c$. We will often abuse this terminology and
say ``$a$ is a stem  in $c$'' when we mean that $c$ contains a stem isomorphic
to $a$. The context should ensure that no confusion arises.
For example, we say that causets $a$ and $b$ ``have the same stems''
when any $d \in \Omega(\N)$ is a stem in $a$ if and only
if it is a stem in $b$.}

Note on terminology: strictly, 
``$a$ is a stem  in $c$'' and ``$c$ contains $a$ as a stem''
mean that $a$ is a subset of $c$. We will often abuse this terminology and
say ``$a$ is a stem  in $c$'' when we mean that $c$ contains a stem isomorphic
to $a$. The context should ensure that no confusion arises.   
For example, we say that causets $a$ and $b$ ``have the same stems'' 
when any $d \in \Omega(\N)$ is a stem in $a$ if and only 
if it is a stem in $b$.

Equally evident is the significance of any question built up as a
logical combination of stem questions of this sort.  To such
compound stem questions belong members of $\R$ built up from stem sets
$\stem(b)$ using union, intersection and complementation (corresponding to
the logical operators `or', `and' and `not', respectively).  If all the members
of $\R$ were of this type, we would not only have succeeded in
characterizing the dynamically meaningful covariant questions at a
formal level, but we would have understood their physical significance
as well.
The following  theorem asserts that, to all intents and
purposes, this is the case. 

\begin{theorem} \label{main.theorem}
For every CSG dynamics as described in section 2,
the family $\cal{S}$ of all stem sets  generates\footnote
{A family ${\cal F}$ of subsets is said to {\it generate} a
 $\sigma$-algebra ${\cal A}$ if ${\cal A}$ is the smallest
 $\sigma$-algebra containing all the members of ${\cal F}$.  
 For example, the cylinder sets introduced above generate the
 $\sigma$-algebra $\Rtilde$.}
the $\sigma$-algebra $\R$ up to sets of measure zero.
\end{theorem}

This is a little vague so let us work towards formulating a 
more precise  statement. Let $\cal{R}({\cal{S}})$ be the $\sigma$-algebra 
generated by $\cal{S}$. Since $\cal{S} \subset \cal{R}$ we know that 
$\cal{R}({\cal{S}}) \subset \cal{R}$.  Unfortunately the latter 
inclusion is strict: there exist sets in $\cal{R}$ which are 
not in $\cal{R}({\cal{S}})$. 
The following is an example. 
Let
\[ 
   \Gamma = \SetOf{c\in\Om}{c \mbox{ contains a maximal element}}
\]
and
\[\Gamma_k^n=\{\ti{c}\in \tOm:\;\; e(k)\not\prec e(m)\;\; \mbox{for}\;
k< m\leq n\}
\]
where $e(j)$ is the element of $\ti{c}$ labeled $j$. 

This latter set is a finite union of cylinder sets, since the condition
requires the initial stretch of $\ti{c}$ 
to be in a particular subset of $\tOm(n)$, 
and
\[
  \Gamma=\overset{\infty}{\underset{k=0}{\cup}}\;
         \overset{\infty}{\underset{n=\kpion}{\cap}}\,
  \Gamma_k^n\] 
Therefore $\Gamma\in\tcR$ and, because the defining
condition of $\Gamma$ is manifestly covariant, $\Gamma$ is moreover
in $\cR$ .

\begin{figure}[ht]
\centering \resizebox{!}{1.5in}{\includegraphics{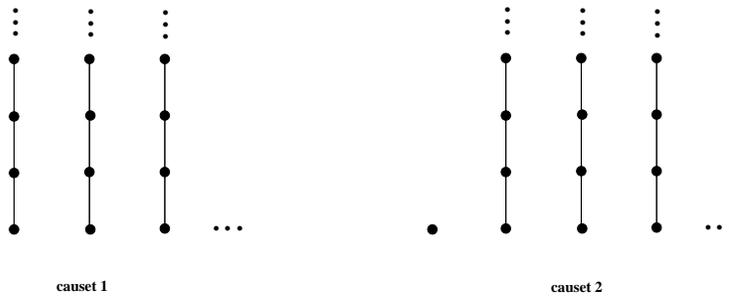}}
\caption{\small{Every stem in causet 1 is also found
in causet 2 and vice versa. 
Since they have the same stems, they cannot be  separated by sets in $\cS$ 
nor in $\cR(\cS)$.}
\label{rogues.fig}}
\end{figure}

To show that $\Gamma\notin\cR(\cS)$, we argue as follows.    
If there exist two completed causets $x,y\in \Om$ such
that every stem set $S\in \cS$ contains either both $x$ and $y$ or 
neither, then the same holds for every $A\in \cR(\cS)$. This
is because
\[
  \cR'\equiv\{ A\in \cR(\cS):\;\; \mbox{either}\;\;
  x,y\in A\;\; \mbox{or}\;\;
  x,y\in A^c\}
\] 
is a $\sigma$-algebra and $\cR'$ contains $\cS$ and 
therefore contains $\cR(\cS)$. Consider now
the following two causets: $c_1$ is the union of infinitely many 
unrelated infinite
chains (a chain is a totally ordered set) 
and $c_2$ is the union of $c_1$ and a single unrelated element   
(see figure \ref{rogues.fig}). Clearly,
$c_1\notin \Gamma$, while $c_2\in \Gamma$.  
Now $c_1$ and $c_2$
cannot be separated by sets in $\cS$: if a  finite causet is a  
stem in $c_1$ it is also a stem in $c_2$ and vice versa.\footnote
{This is in accord with theorem 2 below: if stem sets are not
 generating in the quotient Borel space $(\Om,\cR)$, then they cannot be
 separating either.}
Therefore $\Gamma$, which does separate the two, cannot be in $\cR(\cS)$. 
\vspace{0.5cm}
\vspace{0.5cm}

In this example, the two causets responsible for the failure of $\cal{R}$ and 
$\cal{R}({\cal{S}})$ to be equal have the property that 
they are non-isomorphic but have the same stems.
This suggests that the ``difference'' between the two $\sigma$-algebras
is due to such causets, which we call ``rogue'' causets.
A causet $c\in\Om$ is a rogue if there exists a non-isomorphic 
causet $c'\in \Om$ such that if $b \in \Om(n)$ is 
a stem in $c$ then it is a stem in $c'$ and vice versa. 
Let $\Theta$ be the set of all rogues in $\Om$.

Now we state two propositions that will easily imply our result.  

\begin{proposition}\label{zero.proposition}
$\mu(\Theta) = 0$ in any CSG dynamics
\end{proposition}

\begin{proposition}\label{main.proposition} 
For every set $A\in\cR$, there is
a set $ B\in \cR(\cS)$ such that
$A\triangle B\subset \Theta$.
\end{proposition}
Here $\triangle$ denotes symmetric difference. 

The following immediate corollary is a precise version of 
Theorem~\ref{main.theorem}.  

\begin{corollary}\label{main.corollary}
 For every set $A\in\cR$, there is
a set $ B\in \cR(\cS)$ such that, in any CSG dynamics,
$\mu(A\triangle B) =0$. 
\end{corollary} 

The rest of the paper is devoted to proving the two propositions.

\section{Proof of Proposition \ref{zero.proposition}}

We begin with a little terminology.   
An element $x$ of a causet is {\em maximal} if there is no $y$ with
$x \prec y$ and {\em minimal} if there is no $z$ with $z \prec x$.
The {\em past} of an element $x$ is the set of elements below 
$x$.
A {\em chain} of length $k$ in a causet is a sequence 
$x_0\prec \cdots \prec x_k$.  The {\em level} of an element $x$ in a 
causet is the maximum length of a chain with top element 
$x$ --- so elements at level~0 are exactly minimal elements, and 
those at level~1 are the non-minimal elements that are above only
minimal elements.  As causets are past finite, every element has some 
finite level.  Naturally, {\em level $k$} in a causet consists of 
the elements of level $k$.  If $a$ is a causet, $a_{(k)}$ denotes the
set of all elements of level less than or equal to $k$.

We shall actually characterize exactly the set of rogues, 
although this is more than we need. 

Let $\Gamma = \SetOf{c \in \Omega}{c \hbox{ has a level containing 
infinitely many non-maximal elements}}$. 
Then we have 
\begin{lemma}\label{gammaintheta.lemma}
$\Gamma \subset \Theta$.
\end{lemma}

\bproof
Let $a \in \Gamma$. Suppose level 0 in $a$ has infinitely 
many non-maximal elements. If level 0 in $a$ also contains any non-zero 
number of maximal elements, let causet $b$ be formed from $a$ by deleting 
all those level 0 maximal elements. If level 0 in $a$ 
contains no maximal elements, let $b$ be formed by adding a maximal 
element to level 0. Then $a\not\simeq b$ and they  have the same 
stems. 

Now suppose that level $k$ is the first level in $a$ with infinitely 
many non-maximal elements, so that all levels below $k$ have finitely 
many non-maximal elements.  As $a_{(k-1)}$, the set of elements 
at levels below $k$, is finite, 
there exist an infinite number of non-maximal elements in level 
$k$ of $A$ which all share the same past --- a given subset 
$c\subset a_{(k-1)}$. If there is any non-zero number of maximal elements
in level $k$ of $a$ with past $c$, then let causet $b$ be formed from 
$a$ by deleting all those elements. If there are no maximal elements
in level $k$ of $a$ with past $c$, then let causet $b$ be formed from 
$a$ by adding a maximal element with past $c$.  
Then $a\not\simeq c$ and they have the same
stems.
\eproof

\begin{lemma}\label{thetaingamma.lemma} 
$\Theta \subset \Gamma$.
\end{lemma}

\bproof 
Consider $a \not\in \Gamma$.  Suppose that $b \in \Omega$ has 
the same stems as $a$.  We want to show that $a\simeq b$, which 
then implies that $a \not\in\Theta$.

The plan is to construct partial isomorphisms between $a_{(k)}$
and $b_{(k)}$ for each $k$, and then show that some subsequence
of these partial isomorphisms extends to an isomorphism of the 
completed causet.  Our task is complicated by the possible infinite
sets of maximal elements at each level.  We deal with these by 
a process that we have called ``Hegelianization".

A completed causet $a$ is Hegelianized as follows. First one defines
an equivalence relation in $a$ by setting $x\sim y$ iff either $x=y$
or they are both maximal, have the same past and there are infinitely
many elements with that same past.  The Hegelianization of $a$ is then
the quotient causet $a^*= a/\sim$; since $a\notin \Gamma$, all
levels of $a^*$ are finite.  To each element in $a^*$ that is an 
infinite equivalence class, we attach a {\em flag}: note that 
every unflagged element of $a^*$ is a single-element equivalence 
class $\{ x\}$, which we naturally identify with the element $x$ of
$a$.  

For each $k$, we claim that there is an isomorphism between 
$a^*_{(k)}$ and $b^*_{(k)}$ that preserves the flags.  In showing this, 
we may assume that the following hold:  \newline
(i) $a^*_{(k)}$ contains at least as many flagged elements as 
$b^*_{(k)}$, \newline
(ii) if the two sets contain the same number of flagged elements,  
then $|a^*_{(k)}| \ge |b^*_{(k)}|$. \newline
(Otherwise, exchange the roles of $a$ and $b$.) 

Let $N = |b^*_{(k)}| + 1$.  Let $c$ be obtained from $a^*_{(k)}$
by (a)~replacing each flagged element $x$ by $N$ elements 
$x_1, \dots, x_N$, and (b)~for each element $y$ of $a^*_{(k)}$ 
that is maximal in $a^*_{(k)}$ but not in $a$, include some element 
$z_y$ above $y$ in $a$, and also all elements in the past of $z_y$.  
So $c$ is a stem in $a$, in which all the newly introduced 
elements $x_i$ are maximal.  Hence $c$ is also a stem in $b$, 
i.e.\ there is an embedding $\varphi: c \hookrightarrow b$ whose 
image is a stem in $b$.  Note that $\varphi$ preserves levels.  
So, for any flagged element $x$ of $a^*_{(k)}$, all of 
$\varphi(x_1), \dots, \varphi(x_N)$ are different elements 
of $b_{(k)}$, so by choice of $N$ at least one of them, 
say $\varphi(x_1)$, is in an infinite equivalence class.  
If $y \in a^*_{(k)}$ is not maximal in $a$, then 
$\varphi(y) < \varphi(z_y)$, so $\varphi(y)$ is not maximal in $b$.

We now obtain a natural map 
$\varphi^*: a^*_{(k)} \hookrightarrow b^*_{(k)}$, defined
by taking each unflagged element $z \in a^*_{(k)}$ to 
the equivalence class containing $\varphi(z)$, and each flagged
element $x$ to the equivalence class containing $\varphi(x_1)$. 

We claim that distinct elements $x$ and $y$ of $a^*_{(k)}$ are 
mapped to distinct elements of $b^*_{(k)}$ by $\varphi^*$.   
If neither $x$ nor $y$ is maximal in $a$, then this is immediate 
since $\varphi(x) \not= \varphi(y)$, and these elements are 
unflagged in $b$.  If $x$ is maximal in $a$ and $y$ is not, then 
$\varphi^*(x)$ is maximal in the image of $\varphi^*$ while 
$\varphi^*(y)$ is not.  If $x$ and $y$ are both maximal in $a$, 
but not in the same equivalence class, then either they have 
different pasts --- in which case they are certainly mapped to 
different elements --- or there are only finitely many elements 
with the same past as $x$ and $y$.  In this last case, we have 
to rule out the possibility that the unflagged elements $x$ and 
$y$ are mapped to a flagged element of $b^*_{(k)}$; this is not 
possible, since we have 
shown that distinct flagged elements of $a^*_{(k)}$ are mapped 
to distinct flagged elements of $b^*_{(k)}$, and by assumption 
there are no more flagged elements of $b^*_{(k)}$. 

This shows that $\varphi^*$ is an embedding of $a^*_{(k)}$ 
into $b^*_{(k)}$, and furthermore that the two sets have the
same number of flagged elements.  
As $|a^*_{(k)}| \ge |b^*_{(k)}|$, the map $\varphi^*$ is actually
an isomorphism.  In particular, $|a^*_{(k)}| = |b^*_{(k)}|$ for 
each~$k$.  

Now take a sequence $(\varphi^*_k)$, where $\varphi^*_k$ is
an isomorphism from $a^*_{(k)}$ to $b^*_{(k)}$.  We construct
an isomorphism from $a^*$ to $b^*$ by a standard ``compactness'' 
procedure.  Note that there are only finitely many 1-1 maps 
from level~0 of~$a^*$ to level~0 of~$b^*$; each $\varphi^*_k$
induces one of these maps, so one such map occurs infinitely 
often, say for all $k$ in the infinite set $K_0$.  Similarly, 
there is an infinite subset $K_1$ of $K_0$ such that each 
$\varphi^*_k$ for $k \in K_1$ induces the same map from
level~1 of $a^*$ to level~1 of $b^*$.  We continue in this way
for all levels.  The isomorphism $\Phi$ is then defined by 
combining the 1-1 maps obtained at each level: $\Phi$ is clearly
order-preserving and a bijection. 

Having found an isomorphism from $a^*$ to $b^*$, it is trivial 
to extend it to an isomorphism from $a$ to $b$, as required. 
\eproof

\begin{corollary}
$\Theta = \Gamma$.
\end{corollary}

\begin{lemma}\label{roguesborel.lemma}
 $\Theta$ is measurable.  Indeed $\Theta\in\cR(\cS).$
\end{lemma}
\bproof We show that $\Gamma$ can be constructed 
countably from the stem sets, which are themselves 
measurable.

Let $\Gamma_{kl}^{n_0\cdots n_{k-1}}$ be defined as 
the union of finitely many stem sets as follows. Each stem set in the
union is 
defined by a stem which: 
is a finite causet with
$n_i$ elements in level $i, \; i=0,\cdots, k-1$, $l$ elements in level
$k$ and $m\leq l$ elements in level $k+1$, ordered in such a way that
all the elements in levels $i\leq k$ are non-maximal. There are only
finitely many such stems and the union is over all corresponding stem sets. 

Define $\Gamma_{kl}$ to be the  union of these over  $n_i$, 
$i=0,\cdots, k-1$ for fixed $k$ and $l$:  
\[\Gamma_{kl}=\overset{\infty}{\underset{n_{\kmion}=1}{\cup}}\cdots
\overset{\infty}{\underset{n_0=1}{\cup}}\, \Gamma_k^{n_0\cdots
n_{k-1}} \ .
\]
This is the set of all causets with at least $l$ non-maximal points 
at level $k$. Then, taking the intersection over $l$ and finally the 
union over $k$ we see that
\[
  \Gamma =\,\overset{\infty}{\underset{k=0}{\cup}}
  \,\overset{\infty}{\underset{l=0}{\cap}}
  \,\Gamma_{kl}\ .
\] 
\eproof

\begin{lemma}

{In the CSG dynamics with $t_k \ne 0$ for some
$k > 1$, a causet containing an infinite level almost 
surely does not occur.}
\end{lemma}
\bproof
As we have already seen, a causet contains an infinite 
level if and only it if contains an infinite antichain 
(an antichain is a totally unordered set) all of 
whose elements share the same past.

Fix any labeled stem $\Ctilde$, and let $n_0$ be the largest label
in $\Ctilde$; we show that there are almost 
surely only finitely many elements of the causet with past $\Ctilde$.  
It is enough to show that the expected number of elements with
past equal to $\Ctilde$ is finite (this is exactly the 
Borel-Cantelli Lemma).  The expected number of elements with past 
$\Ctilde$ is the sum, over $n>n_0$, of the probability that $e(n)$, the
element labeled $n$, has past $\Ctilde$.  

Recall that this probability is 
\begin{equation}\label{tau.equation}
\tau_n=\frac{\sum_{l=m}^\varpi \binom{\varpi-m}{l-m}t_l}
              {\sum_{j=0}^n\binom{n}{j}t_j}
\end{equation}
where $\varpi$ is the number of elements in $\Ctilde$ and $m$ the 
number of maximal elements in $\Ctilde$.  Therefore the expected
number of elements with past $\Ctilde$ is 
\[ 
\sum _{n=n_0+1}^\infty \tau_n = 
\sum_{l=m}^\varpi \binom{\varpi-m}{l-m}t_l
\sum _{n=n_0+1}^\infty \frac{1}{\sum_{j=0}^n\binom{n}{j}t_j}.
\]
Because there is some $j \ge 2$ with $t_j >0$, the terms in the last
sum above are bounded above by those in the convergent series
$\sum 1/\binom{n}{j}t_j$.  Hence the sum is finite, and the expectation
is finite, as required.  
\eproof

Since $\Theta$ is a subset of the set of causets with an infinite level
we have:  
\begin{corollary}
$\Theta$ has measure zero.
\end{corollary}
This is our proposition \ref{zero.proposition}.

\section{Some results in measure theory}

In order to prove proposition \ref{main.proposition}
it will be necessary to be a little more
formal than we have been heretofore.  In this section we collect some
relevant definitions and results from measure theory.  For the most part, 
we follow the terminology of Mackey~\cite{Mackey:1976}. 

Recall that a $\sigma$-algebra $\cR$ on a set $X$ is a non-empty family
of subsets of $X$ closed under complementation and countable
union\footnote
{Equivalently, one can require closure under countable intersection
 instead of union.  The two conditions imply each other in the presence
 of closure under complementation.};
that is, $\cR$ must satisfy: 
(i) if $A\in\cR$ then $A^c\in\cR$, and
(ii) if $A_n\in\cR$ for every $A_n$ in a countable family, then 
$\bigcup_n A_n\in\cR$.
By a {\em Borel space} we will mean a pair $(X,\cR)$ where $X$ is a set 
and $\cR$ is a $\sigma$-algebra on $X$.
The members of $\cR$ will be called {\em measurable} subsets
or {\it Borel} subsets.

For any family $\cF$ of subsets of $X$, the $\sigma$-algebra generated
by $\cF$, denoted $\cR(\cF)$,  is the smallest
$\sigma$-algebra that includes every member of $\cF$.  
If $\cF$ is a family of subsets of $X$ and $A$ an arbitrary subset of $X$, 
we write $\cF\cap A$ to denote the family $\SetOf{F\cap A}{F\in\cF}$
of subsets of $A$. 
If $\cF$ is a $\sigma$-algebra, then so also is $\cF\cap A$.
If $(X,\cR)$ is a Borel space and $A$ is an element of $\cR$, then
we call the pair $(A,\cR\cap A)$ a {\em Borel subspace} of $(X,\cR)$.
(We emphasize that according to this definition, only a {\it measurable}
subset $A$ yields a Borel subspace.)

Given two Borel spaces $(X,\cR)$ and $(X',\cR')$, a map $f:X\rar X'$
is said to be a {\it Borel map} if for each $A\in\cR'$, the set
$f^{-1}(A)$ is in $\cR$.  
For any equivalence relation in a Borel
space $(X,\cR)$, a $\sigma$-algebra $\cR'$ is induced in the space 
$X'$ of equivalence classes by requiring the projection $p:X\rar X'$ to
be a Borel map. 
Concretely, $\cR'$ is the family of
subsets $A'\subset X'$ such that $p^{-1}(A')\in\cR$. 
The derived Borel space $(X',\cR')$ is called a {\em quotient} of $(X,\cR)$.  
A quotient of a Borel subspace of a Borel space $(X,\cR)$ is called a
{\em Borel subquotient} of $(X,\cR)$.

\begin{lemma}\label{subobject.lemma} 
  Let $A\subset X$ be a measurable subset in the Borel space $(X,\cR)$.
  Then $\cR(\cF\cap A) = \cR(\cF)\cap A$.
\end{lemma}

This is intuitively obvious because intersecting with
   $A$ preserves complement and countable union.
   A proof is given in \cite{Billingsley:1986}, page 132.

A family $\cF$ of measurable subsets of $X$ is said to {\it separate}
a Borel space $(X,\cR)$ (or to be a {\it separating family} for $(X,\cR)$)
if for every two points $x$, $y\in X$,
there exists a set $U\in\cF$ with $x\in U$ and $y\notin U$.

Naturally associated with any topological space $X$, is the
$\sigma$-algebra generated by the family of open (or equivalently
closed) subsets of $X$.  This is called the {\it topological
$\sigma$-algebra} of $X$.

A topological space $X$ is {\em separable} if it contains a countable
dense subset $D$.  (When $X$ is a metric space, this signifies that every
open ball in $X$ contains a point of $D$.)  
A {\em Polish space} is a separable complete metric space.

We are now in a position to state a key theorem we will use in 
proving Proposition \ref{main.proposition}.
\begin{theorem}\label{mackey.theorem}  
 In a Borel subquotient of a Polish space, 
 any countable separating family is also a generating family.
\end{theorem}

\bproof
Combine the second theorem on page 74 of \cite{Mackey:1976} with the
corollary on page 73, bearing in mind the definition of a standard
Borel space as a Borel subspace of a Polish space.
\eproof

\section{Proof of Proposition \ref{main.proposition}}

We now place the spaces and measures of our causal set stochastic 
process within this formal framework. 
Let $\cC$ be the family of cylinder sets in $\tOm$. 
It is countable, since its members are in one-to-one correspondence with
the collection of finite labeled causets.
Each particular choice of CSG dynamics assigns a probability 
to each set in $\cC$.  
Standard techniques then assure us that this assignment extends in a
unique way to a probability measure $\ti\mu$ in the Borel space
$(\tOm,\tcR)$, 
where as before $\tcR=\cR(\cC)$ is the $\sigma$-algebra generated by $\cC$.

\begin{lemma}\label{polish.lemma}
  There is a metric on $\tOm$ 
  with respect to which $\tOm$ is a Polish space 
  whose topological $\sigma$-algebra is  $\tcR$.
\end{lemma}
\bproof
For each pair of completed labeled causets $\ti{a}$,
$\ti{b} \in \tOm$, we set:
\be\label{metric.eq}d(\ti{a},\ti{b})=1/2^n, \ee 
where $n$ is the largest integer for which $\ti{a}_{(n)}=\ti{b}_{(n)}$. 
It is easy to verify that this gives a metric\footnote
{Indeed the metric $d$ given by (\ref{metric.eq}) satisfies a condition
 stronger than the triangle inequality: for any three causets $\ti{a}$,
 $\ti{b}$ and $\ti{c}$, we have $d(\ti{a},\ti{c}) =
 \max(d(\ti{a},\ti{b}),d(\ti{b},\ti{c}))$.  This `ultrametric' property
 is related to the tree structure of the space $\cC$ of cylinder sets.}
on $\tOm$.  The maximum distance between two causets is $1/2$ and occurs
when their initial two elements already form distinct partial orders.
Note also that the open balls in this metric are exactly the cylinder
sets.

One can readily verify that the metric space $(\tOm,d)$ given by 
(\ref{metric.eq}) is complete. To see that it is separable
we find a countable dense set in $\tOm$.  Associate with each finite
causet $\ti{a}\in\tOm(n)$, the completed causet which results from
adding an infinite chain to the future of the last element in
$\ti{a}$. It is then clear that for any causet $\ti{c}\in\tOm$ such a
`chain-tailed' causet can be found arbitrarily close to $\ti{c}$. 
Therefore $\tOm$ is a Polish space with respect to the metric $d$. 

The family of open balls about the points in this countable dense set is
exactly the family of cylinder sets, and therefore the topological 
$\sigma$-algebra coincides with $\tcR$, as required.  \eproof

Recall that we have defined $\cR$ as the subalgebra of all
label-invariant Borel sets in $\tcR$ and that one can also think of
$\cR$ as a $\sigma$-algebra in the space of unlabeled causets $\Om$.

Consider the equivalence relation of isomorphism on the set $\tOm$ of
labeled causets.  The set of equivalence classes is in 1-1
correspondence with the set $\Om$ of unlabeled completed causets.  Let
$p: \tOm\rar \Om$ be the projection, which assigns to each labeled
causet $\ti{c}$ the class $[\ti{c}]$ of all its possible labelings,
corresponding to the unlabeled causet $c$.  Sets of the form $p^{-1}(A)$
are label-invariant subsets of $\tOm$; to say that this set is in $\tcR$
(or in $\cR$) is equivalent to saying that the corresponding set $A$ of
unlabeled causets is in $\cR$.  Therefore $(\Om,\cR)$ is a quotient of
$(\tOm,\tcR)$.

Furthermore the measure $\ti{\mu}$ on $(\tOm,\tcR)$ restricts to a
measure $\mu$ on $\cR$, which we can define within either interpretation
of $\cR$.  Viewing $\cR$ as a subalgebra of $\tcR$ in $\tOm$, we have
$\mu= \ti{\mu}_{|\cR}$.  While viewing $\cR$ as the quotient of $\tcR$ in
$\Om$, we have $\mu=\ti{\mu}\circ p^{-1}$.

Define $\tOm_0=\tOm \setminus\ti{\Theta}$ and let
$\tcR_0=\tcR\cap\tOm_0$ be the induced $\sigma$-algebra in $\tOm_0$.
Then $(\tOm_0,\tcR_0)$ is a Borel subspace of $(\tOm,\tcR)$
(it follows from Lemma \ref{roguesborel.lemma} that $\tOm_0$ is a Borel 
subset of $\tOm$).  
Similarly let $\Om_0=\Om\setminus\Theta$ and $\cR_0=\cR\cap\Om_0$ 
Then $(\Om_0,\cR_0)$ is a Borel subspace of $(\Om,\cR)$.

\begin{lemma} 
 The Borel space $(\Om_0,\cR_0)$ is a Borel subquotient of $(\tOm,\tcR)$.
\end{lemma}

\bproof 
We have already seen that 
$(\tOm_0,\tcR_0)$ is a Borel subspace of $(\tOm,\tcR)$.
We will show that the Borel space $(\Om_0,\cR_0)$ is a quotient of
$(\tOm_0,\tcR_0)$ by the projection $p$ into isomorphism classes.  
By definition, a set $A$ is in $\cR_0$ if and only if $A\in \cR$ and
$A\subset \Om_0$. Since $(\Om,\cR)$ is the $p$ quotient of $(\tOm,\tcR)$, 
this is equivalent to saying that 
$p^{-1}(A)\in \tcR$ and $p^{-1}(A)\subset \tOm_0$. But this is precisely 
the statement that $p^{-1}(A)$ is in $\tcR_0$. 
\eproof

\begin{lemma}\label{R0analytic.lemma} 
 The countable family $\cS_0=\cS\cap\Om_0$ 
 of (`rogue-free') stem sets 
 separates $\Om_0$. 
\end{lemma}

\bproof 
Consider two distinct causets $x$, $y\in \Om_0$.  There must be a stem
in $x$ but not in $y$ or vice versa, for otherwise $x$ and $y$ would be
in the set $\Theta$ of rogues.  Assume then without loss of generality
that the finite causet $a$ is present as a stem in $x$ but not in
$y$. Then the rogue-free stem set $S_0(a)= S(a)\cap\Om_0$ has 
$x\in S_0(a)$ and $y\notin S_0(a)$ as required.
\eproof

Applying then Theorem \ref{mackey.theorem}, we conclude that $\cS_0$
generates $\cR_0$. We now show that any $A\in \cR$ can be written as
a disjoint union of a set in $\cR(\cS)$ and a set of rogues. Consider
the decomposition of $A$ as a disjoint union $(A\cap\Om_0)\cup (A\cap
\Theta)$. Since the first set is in $\cR_0$, it suffices to establish
that $\cR_0=\cR(\cS)\cap \Om_0$, which is tantamount to the following 
result. 

\begin{lemma}\label{R0-RS.lemma} 
  The Borel space $(\Om_0,\cR_0)$ is a Borel subspace of $(\Om,\cR(\cS))$. 
\end{lemma}

\bproof 
We know that $\cR_0=\cR(\cS_0)$ and $\cS_0=\cS\cap\Om_0$. 
The result follows from Lemma \ref{subobject.lemma}.
\eproof

In particular we have $\cR_0\subset\cR(\cS)$, so that for every
$A\in\cR$, there is a $B\in \cR(\cS)$ such that $A\cap\Om_0=B$.
Hence, $A=B\cup A_r$, with $B\in\cR(\cS)$ and $A_r\subset \Theta$. It
follows that $A\triangle B = A_r\in \Theta$.

This completes the proof of Proposition~\ref{main.proposition}.

\section{Discussion}

We explicitly excluded two cases of CSG models from 
consideration in the body of the paper: namely that where $t_0$ is the
only non-zero coupling constant and that where $t_0$ and $t_1$ only are
non-zero. The former produces an infinite antichain (the dust universe)
and the latter almost surely produces an infinite union of 
trees in which each element has infinitely many children (the infinite infinite tree).
These models are not ``generic''
in the sense of the Rideout-Sorkin paper (since many transition
probabilities vanish).  But nevertheless our theorem covers 
both models 
because they are deterministic, and any deterministic model
trivially satisfies our main theorem.

\bproof 
We want to prove Corollary \ref{main.corollary}
in the case when there exists a causet c such that for any $A\in \cR$,
$\mu(A) = 1$  if $c \in A$ and $\mu(A) =0$  if $c \not\in A$. 
If $c \in A$ we choose $B = \Omega$ and
if $c \not\in A$ we choose $B$ to be the empty set.
And in both cases $\mu(A\triangle B) =0$. 
\eproof

Our theorem doesn't automatically cover the other 
non-generic CSG models, {\it {i.e.}} those not
in the generalized percolation family.  We conjecture that (as for the
generalized percolation family) 
all of these either are deterministic or preclude
infinite antichains, whence the rogues would be of measure zero for
them.  This would extend our results to the general case.

We have concentrated, in this paper, on the consequences of 
general covariance  as it affects the choice of physically 
meaningful questions. In fact another form of general covariance has
already been imposed, in the derivation of the relevant CSG models 
themselves. Rideout and Sorkin constrain the models to those for
which the probability of a cylinder set depends only on the
unlabeled version of the stem which defines it and not on the 
labeling or order of birth.  
That constraint is somewhat analogous to the invariance of the action of 
a gauge field under gauge transformations. However,
 because there is nothing physical about
a cylinder set itself, it might be objected that 
this is an attempt to impose a physical condition on an unphysical 
object.  Might it be possible to consider dynamical models in more generality,
constructing a measure on $\Omega$ and then 
imposing general covariance (and Bell causality for that matter) 
directly on the measure,
in other words to 
find a dynamics that is fundamentally label free? Would that 
lead to a physical measure different to the one we have 
obtained here? We do not know. 

Another question is whether we should treat all isomorphisms (relabelings)
as pure gauge, as we have done here, or
restrict them in some way, to those which affect only a finite subset of
elements or those which can map each point onto only a finite subset of
other points (finite orbits), for example.
One might worry that treating all relabelings as
pure gauge would force us to make the energy vanish and/or lose all
information about its value,  because it is the analog of treating all
diffeomorphisms (diffeos) as pure gauge, 
including asymptotic translations, {\it {etc.}}
However, we  argue against this concern in two ways.

Firstly, in the context of continuum gravity, say, (or even in Special
Relativity with the metric as background), it is not true that we get
conserved quantities from diffeos.  Rather, in a space{\it{time}} setting,
we get them from variations of the fields which are equivalent to a
diffeo near the final boundary, but globally are not induced by any
diffeo whatsoever (cf ``partial diffeos'') \cite{Sorkin:1986ph}.

Secondly, in terms of ``physical observables'' we can ask whether declaring
asymptotic Poincar{\'e} transformations to be pure gauge would not force
the total energy etc to vanish.  The answer is `yes' if we require the
state vector to be annihilated by gauge generators, but this is too
strong.  What we really should do is just limit the ``physical
observables'' to commute with the gauge generators.  This still leaves
exactly what it should, namely the invariant mass and the magnitude of
the ``spin''.  

The equivalent procedure has not been worked out for quantum
measure theory \cite{Sorkin:1995nj} 
but we strongly suspect that it corresponds precisely to
just restricting the measure to diffeo-invariant
sets of histories (without demanding that the measure itself be
invariant -- except in a cosmological setting where the  {\it{entire}}
past is
included, rather than being encapsulated in an initial condition). This
would then be exactly analogous to what we have done in the present
paper for the classical measure.

\section*{Acknowledgments}
 We  are grateful to  Jeremy Butterfield, Chris Isham, Johan Noldus, 
and Ioannis Raptis for helpful discussions. This work was supported in part
by EPSRC grant GR/R20878/01, NSF grant PHY-0098488
and by Goodenough College. FD and RG dedicate their
work on this paper to the memory of their colleague Sonia Stanciu. Sonia was a 
beautiful and brilliant 
young string theorist at Imperial College and had just won 
a prestigious EPSRC Advanced Fellowship when she died of 
cancer earlier this year.

\bibliography{refs}
\bibliographystyle{h-physrev}

\end{document}